\documentclass[11pt]{article}
\usepackage{graphicx}
\usepackage{cite,graphicx,subfigure,rotating,epsfig}

\newcommand{\BABARPubYear}    {05}

\newcommand{\BABARConfNumber} {009}
\newcommand{\SLACPubNumber} {0000}
\newcommand{\LANLNumber} {0507052}

\input ./babarsym

\def\Bztoksksks	{\ensuremath{\Bz \to \KS\KS\KS}}

\def\Bztophiks	{\ensuremath{B^0 \to \phi\KS}}
\def\Bztokspiz	{\ensuremath{B^0 \to \KS\piz}}
\def\Ksto00	{\ensuremath{B^0 \to \pi^0\pi^0}}
\def\ksksks	{\ensuremath{\KS\KS\KS}}

\def\Bflav {\ensuremath{B_{\mbox{flav}}}\xspace}
\def\Btag {\ensuremath{B_{\mbox{tag}}}\xspace}
\def\Bcp {\ensuremath{B_{CP}}\xspace}
\def\cf {\ensuremath{C}}
\def\sf {\ensuremath{S}}
\def\finalscb{\ensuremath{-0.63^{+0.32}_{-0.28}\, (\mbox{\small stat})\pm 0.04
(\mbox{\small syst})} \xspace}
\def\finalccb{\ensuremath{ -0.10  \pm 0.25\, (\mbox{\small stat}) \pm 0.05, (\mbox{\small syst})} \xspace}

\def\finals{\ensuremath{-0.25^{+0.68}_{-0.61} (\mbox{\small stat}) \pm 0.05 (\mbox{\small syst})}}
\def\finalc{\ensuremath{ 0.56^{+0.34}_{-0.43} (\mbox{\small stat}) \pm 0.04 (\mbox{\small syst})}}

\newcommand{\Brec}{\ensuremath{B_{\CP}}}
\newcommand{\Ups}{\ensuremath{\Upsilon}}
\newcommand{\dte}{\ensuremath{\sigma(\deltat)}}
\newcommand{\mmiss}{\ensuremath{m_{\mbox{miss}}}}
\newcommand{\mb}{\ensuremath{m_{B}}}

\long\def\inst#1{\par\nobreak\kern 4pt\nobreak
    {\it #1}\par\vskip 10pt plus 3pt minus 3pt}

\begin{document}
{\pagestyle{empty}

\begin{flushright}
\babar-CONF-\BABARPubYear/\BABARConfNumber \\
SLAC-PUB-\SLACPubNumber \\
hep-ex/\LANLNumber \\
July 2005 \\
\end{flushright}

\par\vskip 5cm

% Title of the paper
\begin{center}
\Large \bf  Measurement of Time-dependent \CP-Violating Asymmetries in 
\Bztoksksks\ Decays

\end{center}
\bigskip

\begin{center}
\large The \babar\ Collaboration\\
\mbox{ }\\
\today
\end{center}
\bigskip \bigskip

% Abstract
%\begin{center}
\large 

%\end{center}
We present preliminary measurements of the \CP\ asymmetry parameters in \Bztoksksks\ decays,
reconstructing two of the \KS into \pipi and one into \ppz. 
In a sample of 227~M \BB\ pairs collected by the \babar\ detector at the \pep2\
\BF\ at SLAC, we find the \CP\ parameters to be
$\sf =~\finals$ and
$\cf =\finalc$.
Combining this result with the previous \babar\ measurement, obtained from 
events with three \KS\ decaying into \pipi{}, we get
\begin{eqnarray*}
\sf &=&\finalscb \\
\cf &=&\finalccb
\end{eqnarray*}

\vfill
\begin{center}
Submitted at the 
International Europhysics Conference On High-Energy Physics (HEP 2005),
7/21---7/27/2005, Lisbon, Portugal
\end{center}

\vspace{1.0cm}
\begin{center}
{\em Stanford Linear Accelerator Center, Stanford University, 
Stanford, CA 94309} \\ \vspace{0.1cm}\hrule\vspace{0.1cm}
Work supported in part by Department of Energy contract DE-AC03-76SF00515.
\end{center}

\newpage
} % end of pagestyle{empty}

% Input author list file (use the same list as for LP for the moment)
\begin{center}
\small

The \babar\ Collaboration,
\bigskip

B.~Aubert,
R.~Barate,
D.~Boutigny,
F.~Couderc,
Y.~Karyotakis,
J.~P.~Lees,
V.~Poireau,
V.~Tisserand,
A.~Zghiche
\inst{Laboratoire de Physique des Particules, F-74941 Annecy-le-Vieux, France }
E.~Grauges
\inst{IFAE, Universitat Autonoma de Barcelona, E-08193 Bellaterra, Barcelona, Spain }
% ...

\end{center}\newpage

% The body of the paper starts here
\section{INTRODUCTION}
\label{sec:Introduction}
In the Standard Model (SM) \CP violation arises from the 
Cabibbo-Kobayashi-Maskawa (CKM) quark mixing matrix~\cite{ckm}.
Decays of \B\ mesons into charmless hadronic final states with three
kaons are dominated by $b \to s\bar s s$
penguin amplitudes, while other SM amplitudes are suppressed by CKM factors~\cite{one}.
Neglecting these CKM-suppressed contributions, the amplitude of time-dependent \CP\ violation for
these channels is proportional to \stwob, where 
$\beta = \arg(-V_{cd}V_{cb}^*/V_{td}V_{tb}^*)$ is the \CP-violating phase difference 
between mixing and decay amplitudes and $V_{ij}$ are the elements of the CKM matrix.
The time-dependent \CP-asymmetry is obtained by measuring the proper time difference
$\deltat = t_{\CP} - t_{\mbox{tag}}$ between
a fully reconstructed neutral $B$ meson (\Bcp) in the final state \ksksks, and a partially reconstructed
recoil $B$ meson (\Btag). The \Btag\ decay provides evidence that it decayed either as \Bz\
or \Bzb\ (flavor tag).
The decay rate ${\mbox{f}}_+({\mbox{f}}_-)$ when the tagging meson is a $\Bz (\Bzb)$
is given by
\begin{eqnarray}
  \label{eqn:td}
  \lefteqn{\mbox{f}_{\pm}(\deltat) \; = \; \frac{e^{-|\deltat|/\tau_{\Bz}}}{4\tau_{\Bz}} \times }\; \\
   && \left[ \: 1 \; \pm \;
    \: S \sin{( \deltamd\deltat)} \mp C \cos{( \deltamd\deltat)} \: \right] \; , \nonumber
\end{eqnarray}
where $\tau_{\Bz}$ is the neutral $B$ meson mean lifetime and
\deltamd is the \Bz--\Bzb oscillation frequency.
The parameters $C$ and $S$ describe the magnitude of \CP\ violation in the decay and
in the interference between decay and mixing, respectively.
The time-dependent \CP -violating asymmetry is defined as
$A_{\CP} \equiv ({\mbox{f}}_+  -  {\mbox{f}}_- )/
({\mbox{f}}_+ + {\mbox{f}}_- )$.

Since at first approximation $b \to s$  decays can be considered as given by
a single amplitude, no direct \CP\ violation is expected ($\cf\sim0$) and 
$\sf \sim -\eta_f\stwob$, where $C$~($S$) is the parameter for direct (mixing-induced)
\CP violation and $\eta_f = +1(-1)$ corresponds to \CP-even (-odd) final states.
In general, a deviation from these expectations might occur without indicating
the presence of physics beyond the Standard Model, since a second (CKM suppressed)
part is present in the decay amplitude. The interference between these two terms can in general 
produce direct \CP\ violation and introduce a nontrivial relation between 
$S$ and $-\eta_f~\stwob$, as a function of the relative phase and size of the 
two amplitudes. It has been noted that for \Bztoksksks{}
(which is a $\eta_f=+1$ state), as
for the golden mode \Bztophiks{}, this suppressed amplitude is a penguin
contribution as well~\cite{ref:gershon,Cheng:2005ug,Furman:2005xp}, so that the ratio of the two terms is
expected to be of the order of $\lambda^2$, where $\lambda= 0.2258 \pm 0.0014$~\cite{fit1} is
the sine of the Cabibbo angle.

The value of \stwob = $0.726 \pm 0.037$ determined from tree-level
$b \to c \bar c s$ decays is in good agreement with the SM expectaion ~\cite{fit1,fit2}.
On the other hand, $b \to s \bar q q$ processes are dominated by one-loop
transitions, and hence may have contributions from diagrams with new heavy
particles. Thus, a sizable deviation of \sf{} from $-\eta_f \stwob$
would be a signal of physics beyond the Standard Model~\cite{grossman}.

Belle and \babar{} Collaboration have already reported a 
measurement of time-dependent \CP-asymmetry 
in \Bztoksksks{}~\cite{Aubert:2005dy}.
In the case of \babar\, the analysis includes only \KS decaying into \pipi.
%%%%%%%%%%%%%%%%%%%%%%%%%%%%%%%%%%%%%%%%%%%%%%%%%
Since this measurement might be limited in precision by the amount of
data one expects to have at the end of \babar{} experiment, the present work has
the main purpose of improving the precision using the same dataset,
but reconstructing one of the \KS{} in the  \ppz\ decay mode.
%%%%%%%%%%%%%%%%%%%%%%%%%%%%%%%%%%%%%%%%%%%%%%%%%
Because of the absence of charged tracks 
originating from the \Bz decay vertex, we use the vertexing technique 
recently developed for \Bztokspiz~\cite{ref:k0spi0prl}.
%This measurement has still large statistical uncertainty.
%The aim of this study is to measure the \CP\ parameters 
%in a data sample which combines the two final states.

\section{THE \babar\ DETECTOR}
\label{sec:babar}
The \babar\ detector is described elsewhere~\cite{Aubert:2001tu}.
The components are a charged-particle
tracking system consisting of a five-layer silicon vertex tracker
(SVT) and a 40-layer drift chamber (DCH) surrounded by a 1.5-T solenoidal
magnet with an instrumented flux return (IFR), an electromagnetic calorimeter
(EMC) comprised of 6580 CsI(Tl) crystals, and a detector of internally reflected
Cherenkov light (DIRC) providing excellent charged $K-\pi$ separation up to
a momentum of 4.5~\gevc relevant for this analysis.

\section{ANALYSIS METHOD}
\label{sec:Analysis}
The \Bztoksksks\ candidate ($B_{\CP}$) is reconstructed combining
three \KS candidates, two of which are reconstructed in the $\KS \to
\pipi$ mode, while the third is reconstructed in the $\KS \to \ppz$ mode.  We reconstruct $\KS \to
\pipi$ candidates from pairs of oppositely charged tracks. The
two-track composites must form a vertex with a \pipi invariant mass
within 11 \mevcc (about 4$\sigma$) of the nominal \KS
mass~\cite{pdg}.  We form $\piz\to\gamma\gamma$ candidates from pairs
of photon candidates in the EMC. Each photon is required to be
isolated from any charged tracks, to carry a minimum energy of 50\mev, and to 
have the expected lateral shower shape.  We reconstruct $\KS \to \ppz$
candidates from \piz pairs which form an invariant mass $480 <
m_{\ppz} < 520$\mevcc.  \Bztoksksks\ candidates are constrained to
originate from the \epem interaction point using a geometric fit,
based on a Kalman Filter~\cite{treefitter}.  We make a requirement on
the consistency of the $\chi^2$ of the fit which retains 93\% of the
signal events, and rejects about 49\% of other \B decays.  We extract
the $\KS\to \pipi$ decay length $L_{\KS}$ and the
invariant mass ($m_{\gamma\gamma}$) from this fit, and require $100 <
m_{\gamma\gamma} < 141${\mevcc} and $L_{\KS}$ greater than $5$ times
its uncertainty.

For each $B$ candidate we compute two kinematic variables, namely the
invariant mass \mb{} and the missing mass $\mmiss = \sqrt{(q_{\epem} -
\tilde{q}_B)^2}$, where $q_{\epem}$ is the four-momentum of the initial
\epem{} system and $\tilde{q}_B$ is the four-momentum of the
\Bztoksksks{} candidate after a mass constraint on the \Bz{} is applied. 
By construction the linear correlation coefficient between
\mmiss{} and \mb{} vanishes.  
This combination of variables shows smaller correlation (0.86\% on 
reconstructed signal Monte Carlo events
and 1.64\% on the final data sample) and a better
background suppression with respect to the equivalent kinematic variables
$\DeltaE$ and $\mes$  used in the 
\babar{} analysis of this mode with all $\KS \to \pipi$ in the final state~\cite{Aubert:2005dy}.
Using simulations of \Bztoksksks{} and $B^0 \to J/\psi  (\to l^+l^-) K^0_S(\to \pi^0\pi^0)$ decays ($l=e,\mu$) and 
reconstructing  $B^0 \to J/\psi (\to l^+l^-) K^0_S(\to \pi^0\pi^0)$ events on data, we determine
the distribution of $\mmiss$ and \mb{} for signal events.
We find the signal resolution for \mb{} to be about 40\mevcc,
the distribution being asymmetric around the maximum, because of
leakage effects in the EMC. The signal
resolution for \mmiss{}, about 6\mevcc, is dominated by the
beam-energy spread.  We select candidates with \mb{} within
150\mevcc of the nominal \Bz{} mass~\cite{pdg}
and with $5.11<\mmiss<5.31\gevcc$. The region
$\mmiss<5.2\gevcc$ is devoid of signal and used for
background characterization.
Most background originates from continuum $\epem\to\qqbar$
($\q=\u,\d,\s,\c$) events, which we suppress using both production
and decay properties.
To exploit the jet-like topology of continuum events, we
use the angle $\theta_T$ between the thrust axis of the
$B_{\CP}$ candidate and the thrust axis formed from the other charged and
neutral particles in the event.
While $|\cos\theta_T|$ is highly peaked near 1 for
$\epem\to\qqbar$ events, it is nearly uniformly distributed for
\BB{} events. We require $|\cos \theta_T|<0.95$.
Moreover, we calculate the ratio $L_{2}/L_{0}$ of two angular moments defined as
$L_j\equiv\sum_i |{\bf p}^*_i| |\cos \theta^*_i|^j$, where ${\bf
  p}^*_i$ is the momentum of particle $i$ in the \epem{} rest frame,
$\theta^*_i$ is the angle between ${\bf p}^*_i$ and the thrust axis of
the \B{} candidate and the sum runs over all reconstructed particles
except for the \B{}-candidate daughters. 
After all selection requirements are applied, the average candidate multiplicity in events
with at least one candidate is approximately $1.67$, coming from 
multiple $\KS \to \ppz$ combinations. In these cases, we select the candidate
with the smallest $\chi^2=\sum_{i}
(m_i-m_{\KS})^2/\sigma^2_{m_i}$, where $m_i$ ($m_{\KS}$) is the measured
(nominal $\KS$) mass and $\sigma_{m_i}$ is the estimated uncertainty on
the mass of the $i$th $\KS$ candidate. In simulated events, this selection criterion gives the right
answer about $81\%$ of the time. The remaining misreconstructed events,
coming from fake $\KS \to \ppz$ candidates, do not affect the determination
of $\Delta t$ and have a small impact on the other variables used in the final fit
(the largest correlation is $\sim 2.5\%$).

Events coming from $b\to c \bar c s$ would reduce any sensitivity to
departures from the Standard Model, as this process is characterized by a Standard-Model 
\CP\ asymmetry ($S \sim \stwob$ and $C \sim0$). We  therefore remove $b\to c \bar c s$ events by
rejecting all candidates with a $\KS\KS$ mass combination within two times
the experimental resolution of the $\chi_{c0}$ mass. The contribution
from $\chi_{c2}$ is found to be negligible.
Combinatorics from other $B$ decays constitute a further source of 
background. We take this into account by adding a component in the 
likelihood fit (see Sec.~\ref{sec:MLfit}), where the shape of each 
likelihood variable is determined from a simulation of inclusive $B$ decays.

For each \Bztoksksks{} candidate we examine the remaining tracks
in the event to determine the decay vertex position
and the flavor of \Btag. Using a neural network based on kinematic and particle
identification information~\cite{sin2bnewbabar} each event is
assigned to one of seven mutually exclusive tagging categories,
designed to combine flavor tags with similar performance and \deltat{}
resolution.  We parameterize the performance of this algorithm in a
data sample ($B_{\rm flav}$) of fully reconstructed $\Bz\to
D^{(*)-}\pip/\rho^+/a_1^+$ decays. The average effective tagging
efficiency obtained from this sample is $Q = \sum_c \epsilon_S^c (1-2w^c)^2=0.299\pm 0.005$,
where $\epsilon_S^c$ and $w^c$ are the efficiencies and mistag
probabilities, respectively, for events tagged in category
$c=1,2,\cdots7$. For the background, the fraction of events
($\epsilon_B^c$) and the asymmetry in the rate of $\Bz$ versus $\Bzb$
tags in each tagging category are extracted from a fit to the data.

The proper-time difference is extracted from the separation of the
\Brec{} and \Btag{} decay vertices. The \Btag{} vertex is
reconstructed inclusively from the remaining charged particles in the
event. To reconstruct the \Brec{} vertex from
the single \KS{} trajectory we exploit the knowledge of the average
interaction point (IP), which is determined on a run-by-run basis from
the spatial distribution of vertices from two-track events.  We
compute \deltat{} and its uncertainty from a geometric fit to the
$\Ups(4S)\to\Bz\Bzb$ system that takes this IP constraint into
account. We further improve the sensitivity to \deltat{} by
imposing a Gaussian constraint on the sum of the two $B$ decay times
($t_{\CP}+t_{\mbox{tag}}$) to be equal to $2\:\tau_{\Bz}$ with an
uncertainty $\sqrt{2}\; \tau_{\Bz}$, which effectively constrains the
two vertices to be near the \Y4S{} line of flight~\cite{ref:k0spi0prl}. We have verified in
a Monte Carlo simulation that this procedure provides an unbiased
estimate of \deltat{}. Details on
the vertexing algorithm can be found in Ref.~\cite{treefitter}. 

The per-event estimate of the uncertainty on \deltat{} reflects the
strong dependence of the \deltat{} resolution on the $\KS$ flight
direction and on the number of SVT layers traversed by the $\KS$ decay
daughters. In about 97\% of the events 
at least one of the two $\KS$ which decay into \pipi have both pion tracks
 reconstructed from at least 4 SVT hits, leading to sufficient resolution
for the time-dependent measurement. The average \deltat{} resolution
in these events is about 1.0 ps. For events which fail this
criterion or for which $\dte>2.5$ ps or
$\deltat>20$ ps, the \deltat{} information is not used.
However, since $C$ can also be extracted from flavor tagging
information alone, these events still contribute to the measurement of $C$.

\section{MAXIMUM LIKELIHOOD FIT}
\label{sec:MLfit}
We extract the results from unbinned maximum-likelihood
fits to the kinematic, event shape $L_2/L_0$,  $\deltat$, 
and flavor tag variables.
We maximize the logarithm of an extended likelihood function
{\small\begin{eqnarray}
  \lefteqn{{\cal L}(S,C,N_S,N_B,N_{B \bar B}f_S,f_B,f_{B\bar B},\vec{\alpha}) \; = \;
    e^{-(N_S+N_B+N_{\BB})} \; \times } \nonumber \\
  & & \prod_{i \in I} \left[ N_S f_S
      \epsilon^{c}_S{\cal P}_S(\vec{x}_i,\vec{y}_i;S,C) +
      N_B f_B \epsilon^{c}_B {\cal P}_B(\vec{x}_i,\vec{y}_i;\vec{\alpha}) +
      N_{\BB} f_{\BB} \epsilon^{c}_{\BB} {\cal P}_{\BB}(\vec{x}_i,\vec{y}_i;\vec{\alpha}) \right] \times \\
 & & \prod_{i \in II} \left[ N_S (1-f_S)
    \epsilon^{c}_S {\cal P}'_S(\vec{x}_i;C) + N_B (1-f_B)
    \epsilon_B^{c} {\cal P}'_B(\vec{x}_i;\vec{\alpha}) +
     N_{\BB} (1-f_{\BB}) \epsilon^{c}_{\BB} {\cal P}'_{\BB}(\vec{x}_i;\vec{\alpha})\right],\nonumber
\end{eqnarray}}
where $I$ ($II$) is the subset of events with (without) \deltat{}
information.
The $N_X$ ($X$ being signal, continuum background, or \BB\ background) 
represent the $X$ component yield, and $f_X$ the fraction of events with
\deltat{} information.
The probabilities ${\cal P}_X$ (${\cal P}'_X$) 
are products of PDFs for each $X$ hypotheses,
evaluated for each event $i$ from the values of
$\vec{x}_i=\{\mb,\mmiss,L_{2}/L_{0},\mbox{tag},\mbox{tagging
  category}\}$ and $\vec{y}_i=\{\deltat,\sigma_{\deltat}\}$.
The  remaining parameters of the fit are denoted by $\vec{a}$.
For the $B$ background events,
the efficiencies and the mistag probabilities $\epsilon^c_{B\bar B}$ and $w^c$,
respectively, for the tagging category $c$, are fixed to the same values of  the 
signal events.
The observables are sufficiently uncorrelated
that we can construct the likelihoods as the products of one-dimensional PDFs.
The PDFs for signal are parameterized from simulations of signal events.
For background PDFs we
determine the functional form from data in the sideband regions of the
other observables where backgrounds dominate.
We include these regions in the fitted sample
and simultaneously extract the parameters
of the background PDFs along with the fit results.
All the parameters of \BB\ background PDFs are determined
using simulated samples of inclusive $B$ decays. All the parameters of
the likelihood that are not determined simultaneously with $S$ and $C$
in the final fit are 
varied according to their uncertainties in order to estimate systematic errors.

The average $\deltaz$ resolution is 
 dominated by the tagging vertex in the event.
Thus, we can characterize the resolution with a much larger
sample of reconstructed $B \to D X$ decays
(\Bflav\ sample), which we use as signal parameterization.
The amplitudes for the $B_{\CP}$ asymmetries and for the \Bflav\
flavor oscillations are reduced by the same factor due to
wrong tags. Both distributions are convoluted with a common $\Delta t$
resolution function (RF). 
The RF is parameterized as the sum of two Gaussians
with a width proportional to the reconstructed $\sigma_{\deltat}$, and
a third Gaussian with a fixed width of
$8$ ps, which accounts for a small
fraction of outlying events~\cite{sin2bnewbabar}. The first two Gaussians have a
non-zero mean, proportional to $\sigma_{\deltat}$, to account for the
small bias in \deltat{} from charm decays on the \Btag{} side.
Backgrounds are accounted for by adding
terms to the likelihood, incorporated with different assumptions
about their $\Delta t$ evolution and resolution function.

The fit procedure was tested with both a parameterized simulation of a large
number of data-sized experiments and a full detector simulation.
The likelihood of our data fit agrees with the likelihoods from
fits to the simulated data.

\section{SYSTEMATIC STUDIES}
\label{sec:Systematics}
We obtain systematic uncertainties in the \CP\ coefficients $S$ and $C$ due to
the parameterization of PDFs for the event yield in signal and background by
varying the parameters within one standard deviation
(evaluated from a fit to Monte Carlo simulated events).
We evaluate the uncertainties associated with
the assumed parameterization of the \deltat\  resolution function for signal
and \BB-background, a possible difference in the efficiency between \Bz\ and \Bzb, and the fixed values
for \deltamd\ and $\tau_{B^0}$ by varying the parameters within one standard deviation
(extracted from a fit to the \Bflav\ sample).
The sum of the two contributions gives the total error associated with the
PDF parameters.
We estimate different uncertainties associated 
with vertexing. The first is obtained by taking
the largest value of $S(C)_{\rm{fit}}-S(C)_{\rm{true}}$ from fits 
to signal Monte Carlo events. 
Here the $S(C)_{\rm{fit}}$ represents the result of the fit to our signal 
Monte Carlo sample, while $S(C)_{\rm{true}}$ represents input values in the 
Monte Carlo generation.
The second uncertainty is from possible SVT layers misalignment.
We assign a systematic uncertainty on our knowledge of 
the beam spot position by shifting the beam position in the simulation
by $\pm 20~\mu m$ in the vertical direction. The sensitivity due to any calibration problems or time-dependent
effects is evaluated by smearing the beam-spot position by an
additional $\pm20\mu m$ in the vertical direction.
We include an additional contribution from the comparison of
the description of the RF between  
BC vertexing and nominal vertexing in the case of 
$B^0 \to J/\psi K^0_S$ events.
We estimate also the errors due to the effect of doubly
CKM-suppressed decays on the tag side~\cite{Long:2003wq}.
We add these contributions in quadrature to obtain the total
systematic uncertainty. The summary is reported in Table ~\ref{tab:syst}.
The largest contribution is related to the knowledge of the PDF parameters.
For reference, we note that this effect produces a
systematic error of $\pm 1.7$ events on the signal yield.
\begin{table}[hbtp]
\begin{center}
\begin{small}
\begin{tabular}{ccccc} \\ \hline\hline
      & $\Delta~S(+)$ & $\Delta~S(-)$ & $\Delta~C(+)$ & $\Delta~C (-)$  \\
\hline

PDF parameters          & 0.046 & 0.039 & 0.029 & 0.027 \\
vertexing method        & 0.012 & 0.012 & 0.025 & 0.025 \\
SVT alignment           & 0.004 & 0.004 & 0.008 & 0.008 \\
beam-spot               & 0.003 & 0.003 & 0.005 & 0.005 \\
data/MC RF              & 0.006 & 0.006 & 0.001 & 0.001 \\
doubly-CKM-suppressed decays   & 0.001 & 0.001 & 0.011 & 0.011 \\
\hline
total errors            & 0.049 & 0.041 & 0.041& 0.039\\
\hline\hline
\end{tabular}
\end{small}
\caption{Summary of systematic uncertainties on S and C.\label{tab:syst}}
\end{center}
\end{table}

\section{SUMMARY OF RESULTS}
\label{sec:Physics}
\begin{figure}[ht]
\begin{center}
\begin{tabular}{cc}
\epsfig{file=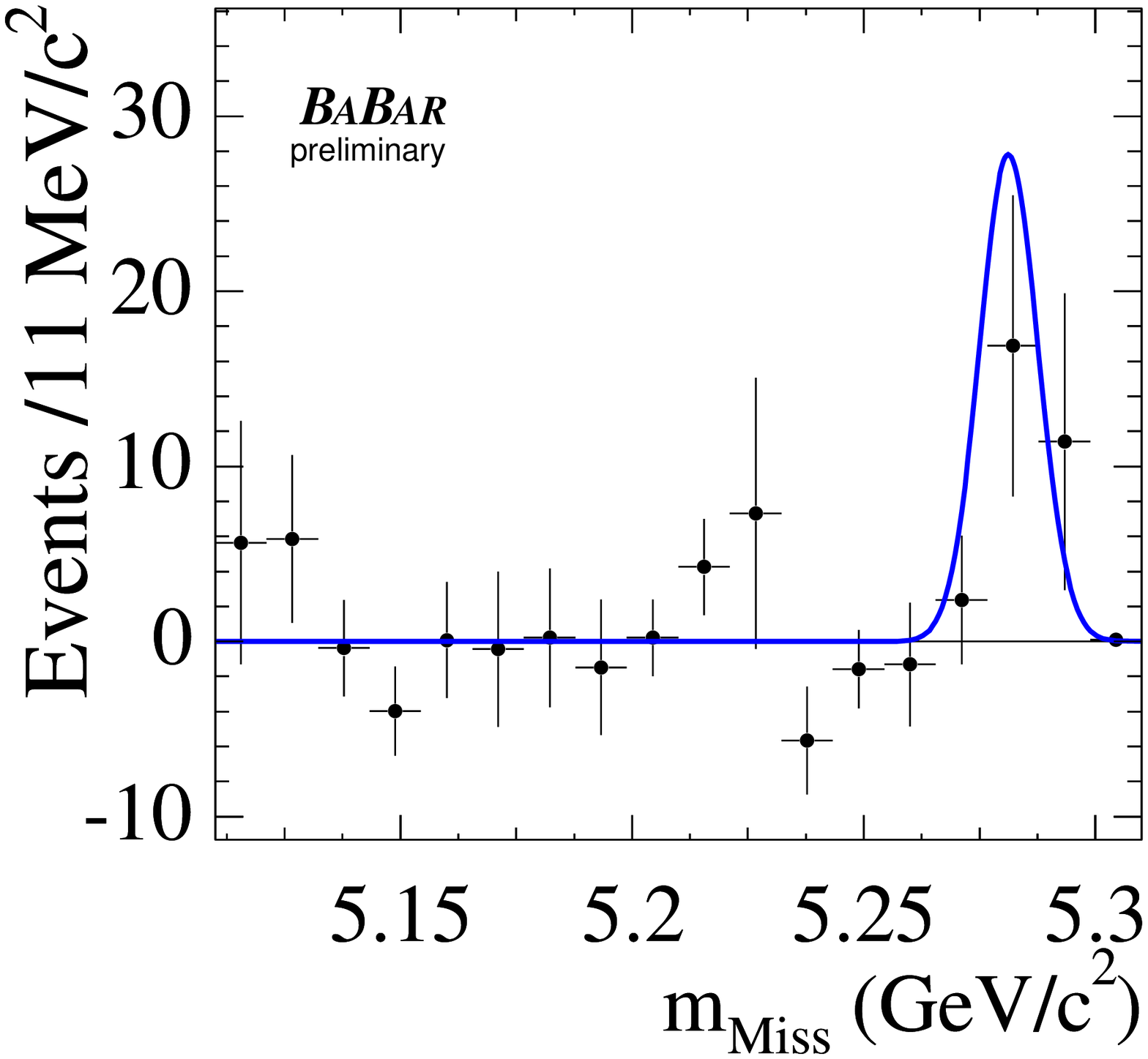,width=8.cm}
\epsfig{file=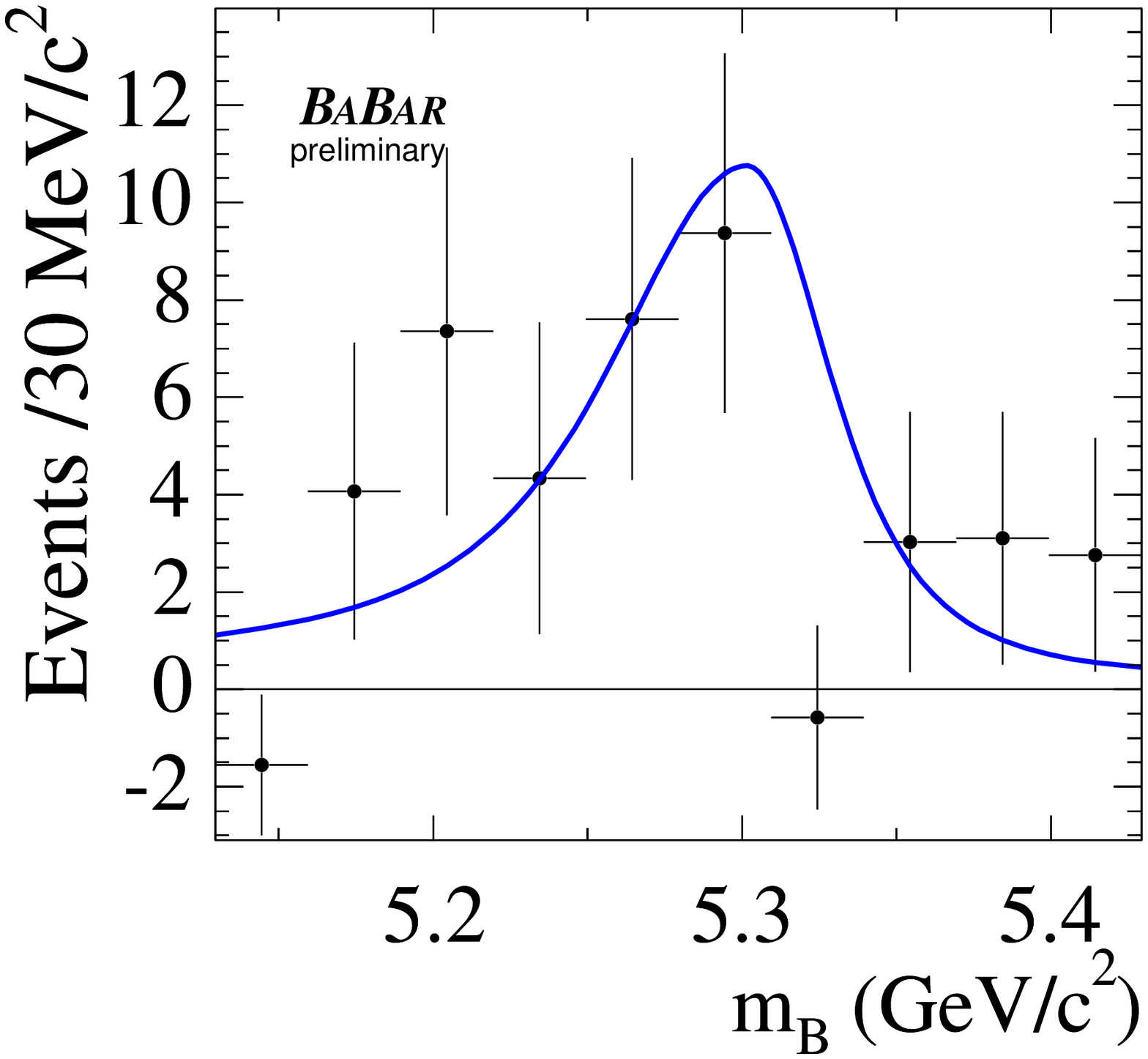,width=8.cm}
\end{tabular}
\caption{\label{fig:splots} Distribution of the event variable \mmiss\ (left)
and \mb\ after reconstruction with the weighting technique described in the text.}
\end{center}
\end{figure}

\begin{figure}[ht]
\begin{center}
\begin{tabular}{cc}
\epsfig{file=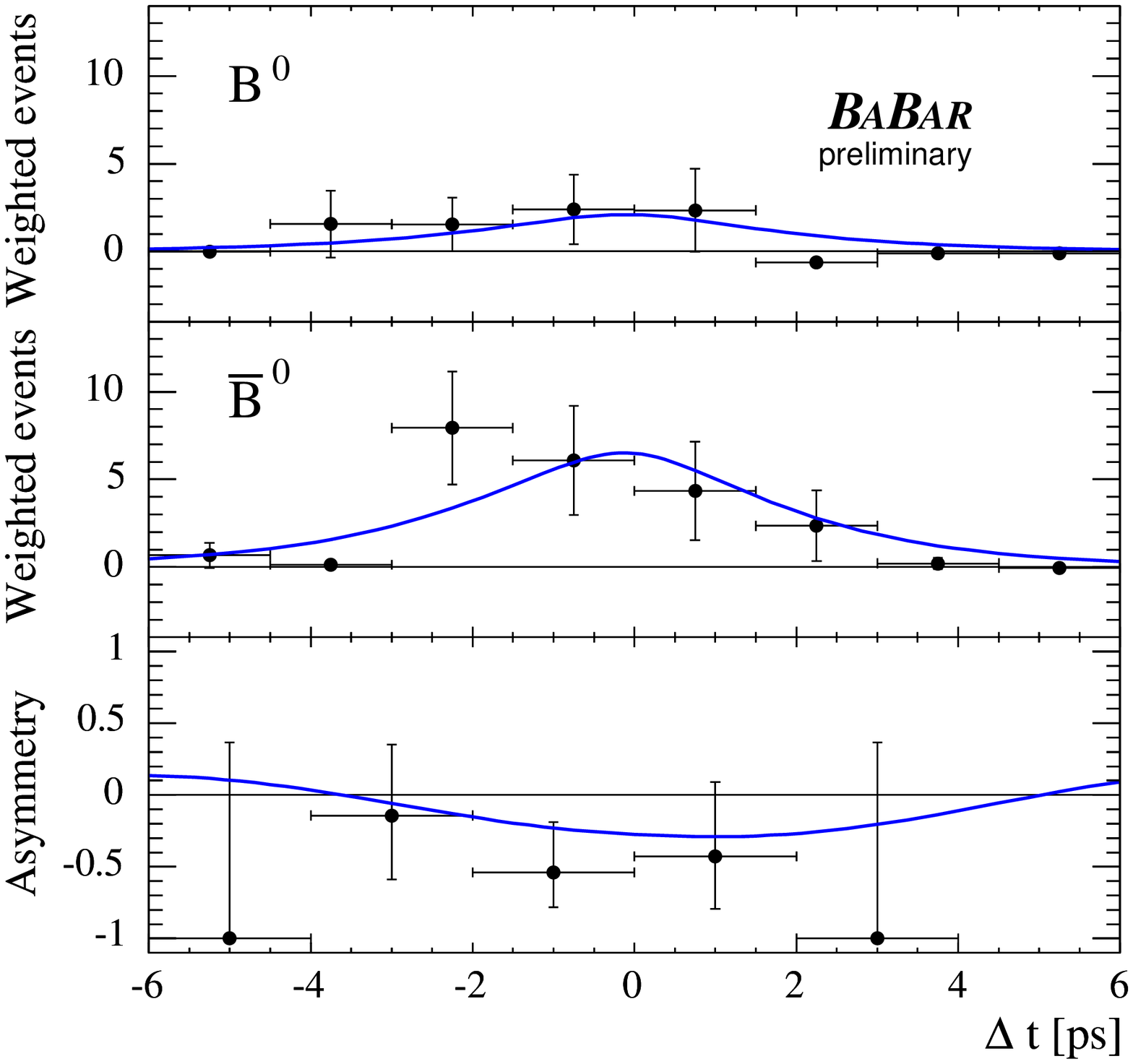,width=8.cm}
\epsfig{file=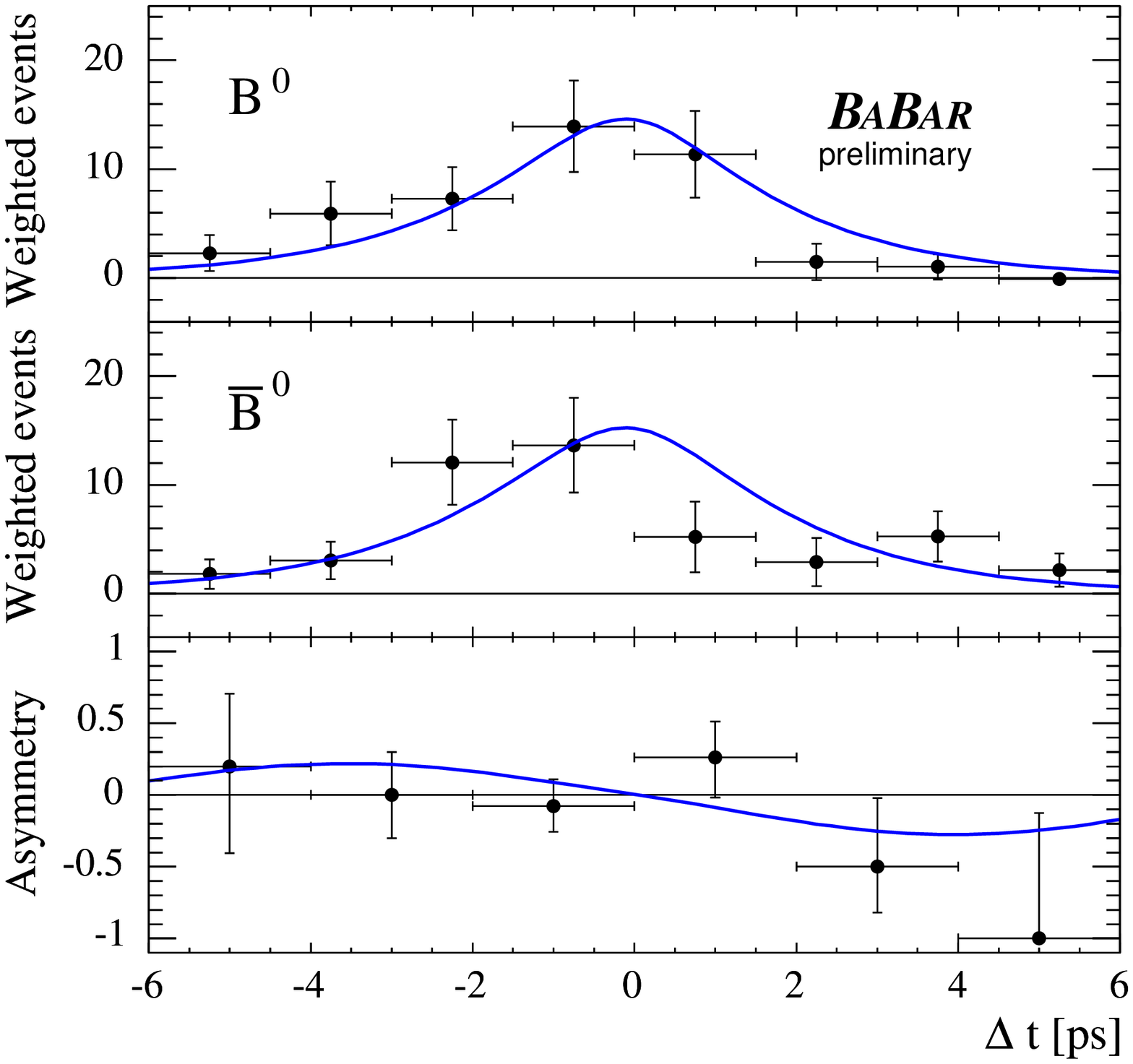,width=8.cm}
\end{tabular}
\caption{\label{fig:dt}Distributions of \deltat{} for weighted events with \Btag{}
tagged as \Bz{} (upper plots) or \Bzb (middle plots), and the asymmetry (lower plots).
Left plots are for the subsample with all $\KS \to \ppz$, right plots are for the 
combined fit. The points are weighted data and the curves are the PDF projections.}
\end{center}
\end{figure}

In the final sample composed of 2748 \B candidates we measure
$41.0^{+9.2}_{-8.3}$ signal events, $2700 \pm 56$ continuum background events
and $7^{+24}_{-19}$ \BB-background events. 
Assuming the world average for the branching ratio ($(6.2 \pm 0.9)10^{-6}$)\cite{Aubert:2005dy} 
and the reconstruction efficiency as estimated from a sample of simulated signal events, 
we expected $45 \pm 7$ events, which is in good agreement with the result.
We find this preliminary result on \CP\ parameters:
\begin{eqnarray*}
S &=&\finals \\
C &=&~ ~\finalc .
\end{eqnarray*}
Fig.~\ref{fig:splots} shows the background-subtracted distributions of \mmiss\ and \mb\
for these events, obtained using the sPlot weighting 
technique~\cite{Pivk:2004ty}. 
Events contribute according to
a weight constructed from the covariance matrix for the yields ($N_S$
and $N_B$) and the probability ${\cal P}_S$ and ${\cal P}_B$ for the
event, computed without the use of the variable that is being
displayed. The curves represent the signal PDFs used in the fit.
We combined this result with the previous \babar\ measurement, obtained 
using \Bztoksksks\ reconstructed from all three \KS decaying 
into \pipi~\cite{Aubert:2005dy}. The combination is obtained through
a simultaneous maximum likelihood fit, which takes into account the correlations
from the common $\Delta t$ PDF. The total systematic error is calculated by summing
in quadrature the uncorrelated sources of errors and taking the largest 
contribution from the two analyses in the case of common sources of background.
In this way, we obtain this preliminary result:
\begin{eqnarray*}
\sf &=&\finalscb \\
\cf &=&\finalccb
\end{eqnarray*}
In Fig.~\ref{fig:dt} we show the distributions of signal events, 
obtained using the sPlot weighting technique~\cite{Pivk:2004ty}, in the case of
\Bztoksksks\ with one \KS reconstructed by \ppz mode (left) and for the combined fit (right). 
The superimposed curves represent the results of the fit in the two cases.

Considering the present uncertainty, this result agrees with Standard Model expectations.
A future update of this analysis, including new data collected by \babar{}, will help
to understand if the present hint of pattern in the deviation of $b \to s$ penguins
from the Standard Model predictions~\cite{Group(HFAG):2005rb} 
is a statistical effect or a signal of new physics.

\section{ACKNOWLEDGMENTS}
\label{sec:Acknowledgments}

% Standard acknowledgments paragraph; must always be included.
We are grateful for the 
extraordinary contributions of our \pep2\ colleagues in
achieving the excellent luminosity and machine conditions
that have made this work possible.
The success of this project also relies critically on the 
expertise and dedication of the computing organizations that 
support \babar.
The collaborating institutions wish to thank 
SLAC for its support and the kind hospitality extended to them. 
This work is supported by the
US Department of Energy
and National Science Foundation, the
Natural Sciences and Engineering Research Council (Canada),
Institute of High Energy Physics (China), the
Commissariat \`a l'Energie Atomique and
Institut National de Physique Nucl\'eaire et de Physique des Particules
(France), the
Bundesministerium f\"ur Bildung und Forschung and
Deutsche Forschungsgemeinschaft
(Germany), the
Istituto Nazionale di Fisica Nucleare (Italy),
the Foundation for Fundamental Research on Matter (The Netherlands),
the Research Council of Norway, the
Ministry of Science and Technology of the Russian Federation, and the
Particle Physics and Astronomy Research Council (United Kingdom). 
Individuals have received support from 
CONACyT (Mexico),
the A. P. Sloan Foundation, 
the Research Corporation,
and the Alexander von Humboldt Foundation.

\end{document}